\documentclass[epj,final]{svjour}

\usepackage{graphicx}
\begin{document}
\title{Coupling and Dissociation in Artificial Molecules}
\author{Constantine Yannouleas\thanks{\emph{email: constantine.yannouleas@physics.gatech.edu}}
\and Uzi Landman 
}
%
%
\institute{School of Physics,
Georgia Institute of Technology, Atlanta, GA 30332-0430}
\date{Received: December 19, 2000 / Revised version: date}
%
\abstract{%
We show that the spin-and-space unrestricted Hartree-Fock method, in 
conjunction with the companion step of the restoration of spin and space 
symmetries via Projection Techniques (when such symmetries are broken), is 
able to describe the full range of couplings in two-dimensional double quantum
dots, from the strong-coupling regime exhibiting delocalized molecular 
orbitals to the weak-coupling and dissociation regimes associated with a 
Generalized Valence Bond combination of atomic-type orbitals localized on the 
individual dots. The weak-coupling regime is always accompanied by an 
antiferromagnetic ordering of the spins of the individual dots. The cases of 
dihydrogen (H$_2$, $2e$) and dilithium (Li$_2$, $6e$) quantum dot molecules 
are discussed in detail. 
} 
\maketitle

\section{Introduction}
\label{intro}
Two-dimensional (2D) Quantum Dots (QD's) are usually referred to as artificial
atoms, a term suggestive of strong similarities between these manmade devices 
and the physical behavior of natural atoms. As a result, in the last few 
years, an intensive theoretical 
effort\cite{cha,pfa,vig,win,leb,man,yl1,yl2,yl3,bar} 
has been devoted towards the elucidation of the appropriate analogies and/or 
differences. Recently, we have shown \cite{yl1,yl2,yl3}
that, even in the absence of a magnetic field, the most promising analogies 
are found mainly outside the confines of the central-field approximation
underlying the Independent-Particle Model (IPM) and the ensuing physical
picture of electronic shells and the Aufbau Principle. Indeed, as a result of
the lower electronic densities in QD's, strong $e-e$ correlations can lead 
(as a function of the ratio $R_W$ between the interelectron repulsion and the
zero-point kinetic energy) to a drastically different physical regime, 
where the electrons become localized, arranging themselves in concentric 
geometric shells and forming electron molecules. In this context, 
it was found \cite{yl2,yl3} that the proper analogy 
for the particular case of a 2e QD is the collective-motion picture
reminiscent of the fleeting and rather exotic phenomena of the doubly-excited
natural helium atom, where the emergence of a ``floppy'' trimeric molecule
(consisting of the two localized electrons and the heavy $\alpha$-particle
nucleus) has been well established \cite{kel,ber}.

A natural extension of this theoretical effort has also developed in the
direction of lateral 2D QD Molecules (QDM's, often referred to as artificial
molecules), aiming at elucidating \cite{yl1,bar,nag,wen,burk,husa}
the analogies and differences between
such artificially fabricated nanostrustures and the natural molecules.

In this paper, we address the interplay of coupling
and dissociation in QDM's, and its effects concerning the appearance of
ferromagnetic versus antiferromagnetic ordering. We will show that this
interplay relates directly to the nature of the coupling in the 
artificial molecules, and in particular to the question whether such coupling 
can be descibed by the Molecular Orbital (MO) Theory or the Valence Bond (VB)
Theory in analogy with the chemical bond in natural molecules.

Furthermore, we show that the onset at a moderate interdot barrier
or interdot distance $d_0$, as well as the permanency for all separations 
$d > d_0$, of spontaneous magnetization and ferromagnetic 
ordering predicted for Double QD's by local-spin-density (LSD) calculations 
\cite{nag,wen} is an artifact of the MO structure implicit in the framework of
these density-functional calculations.

We utilize a self-consistent-field theory which can go beyond the
MO approximation, namely the spin-and-space unrestricted Hartree-Fock 
(sS-UHF), which was introduced by us \cite{yl1,yl2} for the 
description of the many-body problem of both single \cite{yl1,yl2} and 
molecular \cite{yl1} QD's. This sS-UHF employs $N$ (where $N$ is the number of
electrons) orbital-dependent, effective (mean-field) potentials and it
differs from the more familiar \cite{hf} restricted HF (RHF) in two ways: (i)
{\it it relaxes the double-occupancy requirement\/}, namely, it employs 
different spatial orbitals for the two different (i.e., the up and down) spin
directions [DODS, thus the designation ``spin (s) unresricted''], and (ii) 
{\it it relaxes the requirement that the electron orbitals be constrained by
the symmetry of the external confining field\/} [thus the designation ``space
(S) unrestricted''].

In this paper we will show that the solutions with broken 
space symmetry allowed in QDM's by the sS-UHF provide a natural vehicle for 
formulating a Generalized Valence Bond (GVB) theory. Furthermore, they result 
in an antiferromagnetic ordering of the molecular ground state, in
contrast to the ferromagnetic ordering of the MO method which is
associated with HF solutions that preserve the space symmetry, namely 
those derived from the fully restricted Hartree-Fock (RHF) or the spin 
unrestricted (but not space unrestricted) Hartree-Fock (s-UHF) approaches.
     
\section{The two-center-oscillator confining potential}
\label{sec:2}

In the 2D two-center-oscillator (TCO), the single-particle levels 
associated with the confining potential of the artificial molecule are 
determined by the single-particle hamiltonian \cite{note1}
\begin{eqnarray}
H=T &+& \frac{1}{2} m^* \omega^2_{x k} x^2
    + \frac{1}{2} m^* \omega^2_{y k} y^{\prime 2}_k \nonumber \\
    &+& V_{neck}(y) +h_k+ \frac{g^* \mu_B}{\hbar} {\bf B \cdot S}~,
\label{hsp}
\end{eqnarray}
where $y_k^\prime=y-y_k$ with $k=1$ for $y<0$ (left) and $k=2$ for $y>0$ 
(right), and the $h_k$'s control the relative well-depth, thus allowing studies
of hetero-QDM's.
$x$ denotes the coordinate perpendicular to the interdot axis ($y$). 
$T=({\bf p}-e{\bf A}/c)^2/2m^*$, with ${\bf A}=0.5(-By,Bx,0)$, and the last 
term in Eq. (\ref{hsp}) is the Zeeman interaction with $g^*$ being the 
effective $g$ factor and $\mu_B$ the Bohr magneton. 
Here we limit ourselves to systems with $\hbar \omega_{x1}=\hbar \omega_{x2}=
\hbar \omega_x$. The most general shapes described by $H$ are two 
semiellipses connected by a smooth neck [$V_{neck}(y)$]. $y_1 < 0$ 
and $y_2 > 0$ are the centers of these semiellipses, $d=y_2-y_1$ is the 
interdot distance, and $m^*$ is the effective electron mass.

For the smooth neck, we use 
$V_{neck}(y) = \frac{1}{2} m^* \omega^2_{y k} 
[c_k y^{\prime 3}_k + d_k y^{\prime 4}_k] \theta(|y|-|y_k|)$, 
where $\theta(u)=0$ for $u>0$ and $\theta(u)=1$ for $u<0$.
The four constants $c_k$ and $d_k$ can be expressed via two parameters,
as follows: $(-1)^k c_k= (2-4\epsilon_k^b)/y_k$ and
$d_k=(1-3\epsilon_k^b)/y_k^2$, 
where the barrier-control parameters $\epsilon_k^b=(V_b-h_k)/V_{0k}$ 
are related to the actual (controlable) height 
of the bare barrier ($V_b$) between the two QD's, and 
$V_{0k}=m^* \omega_{y k}^2 y_k^2/2$ (for $h_1=h_2$, $V_{01}=V_{02}=V_0$).

The single-particle levels of $H,$
including an external perpendicular magnetic field $B$, 
are obtained by numerical diagonalization in a (variable-with-separation) 
basis consisting of the 
eigenstates of the auxiliary hamiltonian:
\begin{equation}
H_0=\frac{{\bf p}^2}{2m^*} + \frac{1}{2} m^* 
    \omega_x^2 x^2
      + \frac{1}{2} m^* \omega_{yk}^2 y_k^{\prime 2}+h_k~.
\label{h0}
\end{equation}
This eigenvalue problem is separable in $x$ and $y$, i.e., the
wave functions are written as $\Phi_{m \nu} (x,y)= X_m (x) Y_\nu (y)$.
The solutions for $X_m (x)$ are those of a one-dimensional
oscillator, and for $Y_\nu (y)$ they can be expressed through the parabolic
cylinder functions \cite{note1} $U[\alpha_k, (-1)^k \xi_k]$, where
$\xi_k = y^\prime_k \sqrt{2m^* \omega_{yk}/\hbar}$, 
$\alpha_k=(-E_y+h_k)/(\hbar \omega_{yk})$, 
and $E_y=(\nu+0.5)\hbar \omega_{y1} + h_1$ denotes
the $y$-eigenvalues. The matching conditions at $y=0$ for the left and
right domains yield the $y$-eigenvalues and the eigenfunctions
$Y_\nu (y)$ ($m$ is integer and $\nu$ is in general real). 

In this paper, we will focus on the zero-field case ($B=0$) and
we will limit ourselves to symmetric (homopolar) QDM's,
i.e., $\hbar \omega_{x}=\hbar \omega_{y1}=\hbar \omega_{y2}=
\hbar \omega_0$, with equal well-depths of the left and right dots, i.e.,
$h_1=h_2=0$. In all cases, we will use $\hbar \omega_0 =5$ meV and
$m^*=0.067 m_e$ (this effective-mass value corresponds to GaAs).

\section{The Many-Body Hamiltonian}
\label{sec:3}
The many-body hamiltonian ${\cal H}$ for a dimeric QDM comprising $N$ 
electrons can be expressed as a sum of the single-particle part $H(i)$ defined
in Eq.\ (\ref{hsp}) and the two-particle interelectron Coulomb repulsion,
\begin{equation}
{\cal H}=\sum_{i=1}^{N} H(i) +
\sum_{i=1}^{N} \sum_{j>i}^{N} \frac{e^2}{\kappa r_{ij}}~,
\label{mbh}
\end{equation}
where $\kappa$ is the dielectric constant and $r_{ij}$ denotes
the relative distance between the $i$ and $j$ electrons.

As we mentioned in the introduction, we will use the sS-UHF  
method for determining at a first level an approximate solution of the 
many-body problem specified by the hamiltonian (\ref{mbh}). 
The sS-UHF equations were solved in the Pople-Nesbet-Roothaan
formalism \cite{solu} using the interdot-distance adjustable basis formed with
the eigenfunctions $\Phi_{m \nu} (x,y)$ of the TCO defined in section 2.

As we will explicitly illustrate in section 5 for the case of the H$_2$-QDM,
the next step in improving the sS-UHF solution involves the use of 
Projection Techniques in relation to the UHF single Slater determinant.

\begin{figure}[t]
\vspace{3cm}
\centering{\Large{\bf FIGURE 1}}
\vspace{3cm}
\caption{
Lateral Li$_2$-QDM: s-UHF spin-up occupied orbitals (modulus square) and 
total charge (CD) and spin (SD) densities (top row) for the $P=2$ spin 
polarized case. 
The numbers displayed with each orbital are their s-UHF 
eigenenergies in meV. The two spin-down orbitals are not
displayed; they are similar to the $\sigma_g(\uparrow)$ and the 
$\sigma_u(\uparrow)$ and have energies of 17.10 meV and 17.27 meV, 
respectively. The number displayed with the total charge density (top left0 is
the s-UHF total energy in meV. Distances are in nm and the electron densities 
in $10^{-4}$ nm$^{-2}$. The choice of parameters is: $m^*=0.067 m_e$, 
$\hbar \omega_0=5$ meV, $d=70$ nm, $V_b=10$ meV, $\kappa=20$.
}
\end{figure}
\begin{figure}[t]
\vspace{3cm}
\centering{\Large{\bf FIGURE 2}}
\vspace{3cm} 
\caption{
Lateral Li$_2$ QDM: RHF occupied orbitals (modulus square) and charge (CD) and 
spin (SD) densities (top row) for the $P=0$ spin unpolarized case. The numbers
displayed with each orbital are their RHF eigenenergies in meV, while the up 
and down arrows indicate electrons with up or down spin, respectively. The 
number displayed with the total charge density is the RHF total energy in meV.
Distances are in nm and the electron densities in $10^{-4}$ nm$^{-2}$.
The choice of parameters is the same as in Fig.\ 1.
}
\end{figure}
\section{A first example of a homopolar two-dimensional artificial molecule: 
the lateral Li$_2$-QDM}
\label{sec:4}

As an illustrative example, we consider here an open-shell ($N=6$ electrons) 
QDM made of two QD's (hence the name Li$_2$-QDM), with an interdot distance of
$d=70$ nm and an interdot barrier height
of $V_b=10$ meV. The value of the dielectric constant is first taken to be 
$\kappa=20.0$. This value is higher than the value $\kappa$(GaAs)$=12.9$ for
GaAs and may be viewed as resulting from screening produced by the external 
charges residing in the gates and/or the finite height (in the $z$ direction) 
of the dot. This value of $\kappa$ is chosen such that 
the $e-e$ repulsion is not strong enough to precipitate individual electron 
localization \cite{note12} within each individual dot 
(that is, to inhibit Wigner crystallization on each of the dots,
see Ref.\ \cite{yl1}). This example thus conforms better to the case of a 
natural Li$_2$ molecule. It was earlier presented briefly in Ref.\ \cite{yl1} 
in connection with the formation of electron puddles (EP's). Here we will 
present a detailed study of it by contrasting the descriptions resulting from 
both the space-symmetry preserving (RHF or s-UHF) and non-preserving
(sS-UHF) methods. 

We start by presenting HF results which preserve the space symmetry of
the QDM; the results corresponding to the $P=N\uparrow-N\downarrow=2$ and
the $P=0$ spin polarizations are displayed in Fig.\ 1 and Fig.\ 2, 
respectively. The $P=2$ case is an open-shell case and is treated within the 
(spin unrestricted) s-UHF method, in analogy with the standard practice for 
open-shell configurations in Quantum Chemistry \cite{solu2}; the $P=0$ case is
of a closed-shell-type and is treated within the RHF approach.

The corresponding energies are $E_{sUHF}(P=2)=74.137$ meV and
$E_{RHF}(P=0)=75.515$ meV, i.e., the space-symmetry preserving HF 
variants predict that the polarized 
state is the ground state of the molecule. Indeed, by preserving the 
symmetries of the confining potential, the associated orbitals are clearly 
of the MO-type. In this MO picture the Li$_2$-QDM exhibits an open shell
structure (for 2D QDM's, the closed shells correspond to $N=4, 12, 24,...$
electrons, that is shell closures at twice the values corresponding to an
individual harmonic 2D QD) and thus Hund's first rule should apply for the two
electrons outside the core closed shell of the first 4 electrons. 
  
Fig.\ 1 displays for the $P=2$ spin polarization the four occupied spin-up 
s-UHF orbitals ($\sigma_g$, $\sigma_u$, $\pi_{x,g}$, $\pi_{x,u}$),
as well as (top of Fig.\ 1) the total charge density (CD, sum of the
$N\uparrow+N\downarrow$ electron densities) and spin density (SD, difference 
of the $N\uparrow-N\downarrow$ electron densities). The two occupied 
spin-down orbitals ($\sigma_g^\prime$, $\sigma_u^\prime$) are not shown: in
conforming with the DODS approximation, they are slightly different from
the $\sigma_g(\uparrow)$ and $\sigma_u(\uparrow)$ ones. 

Fig.\ 2 displays the corresponding quantities for the $P=0$ case. In both 
(the $P=2$ and $P=0$) cases, the orbitals
are clearly molecular orbitals delocalized over the whole QDM. The $P=2$ 
polarization (Fig.\ 1) exhibits the molecular configuration 
$\sigma_g^1 \sigma_g^{\prime 1} \sigma_u^1 \sigma_u^{\prime 1} 
\pi_{x,g}^1 \pi_{x,u}^1$; this configuration changes to 
$\sigma_g^2 \sigma_u^2 \pi_{y,g}^2$ in the $P=0$ case (Fig.\ 2).

\begin{figure}[t]
\vspace{3cm}
\centering{\Large{\bf FIGURE 3}}
\vspace{3cm} 
\caption{
Lateral Li$_2$-QDM: sS-UHF occupied orbitals (modulus square) and charge (CD) 
and spin (SD) densities (top row) for the $P=0$ spin unpolarized case. The 
numbers displayed with each orbital are their sS-UHF eigenenergies in meV, 
while the up and down arrows indicate an electron with an up or down spin. The 
number displayed with the charge density is the sS-UHF total energy in meV.
Distances are in nm and the electron densities in $10^{-4}$ nm$^{-2}$.
The choice of parameters is the same as in Fig.\ 1.
}
\end{figure}
Fig.\ 3 displays the corresponding quantities for the $P=0$ state calculated 
using the (spin-and-space unrestricted) sS-UHF. This state exhibits a breaking
of space symmetry (the reflection symmetry between the left and right dot).
Unlike the MO's of the $P=0$ RHF solution (Fig.\ 2), the sS-UHF 
orbitals in Fig.\ 3 are well localized on either one of the two individual 
QD's and strongly resemble the atomic orbitals (AO's) of an individual Li-QD,
i.e., they are of $1s$ and $1p_x$ type. Comparing with the MO case of Fig.\ 2,
one sees that the symmetry breaking did not greatly influence the total charge
density. However, a dramatic change appeared regarding the spin densities.
Indeed the SD of the sS-UHF solution (see top right of Fig.\ 3) exhibits a 
prominent spin density wave (SDW) associated with an antiferromagnetic 
ordering of the coupled individual QD's. Formation of such SDW ground states 
in QDM's is accompanied by electron (orbital) localization on the individual 
dots, and thus in Ref.\ \cite{yl1} we proposed for them the name of Electron 
Puddles (EP's). Notice that the EP's represent a separate class of symmetry
broken solutions, different \cite{yl1} from the SDW's that can develop within 
a single QD \cite{mann} and whose formation involves the relative rotation of 
delocalized open-shell orbitals, instead of electron localization.

The sS-UHF total energy for the $P=0$ unpolarized case is
$E_{sSUHF}(P=0)=74.136$ meV, i.e., the symmetry breaking produces a remarkable
gain in energy of 1.379 meV. As a result, the unpolarized state is the
ground state, while the ferromagnetic ordering predicted by the RHF is 
revealed to be simply an
artifact of the MO structure implicit in this level of approximation. Notice 
that the symmetry-broken unpolarized state is only 0.001 meV lower 
in energy than the $P=2$ polarized one, namely the two states are practically
degenerate, which implies that for the set of parameters employed
here the QD molecule is located well in the dissociation regime. 

Although the symmetry breaking within the HF theory 
allows one to correct for the artifact of
the spontaneous polarization exhibited by the MO approaches, i.e., the
the fact that the MO polarized solutions are the lowest in energy upon
separation, the resulting sS-UHF many-body wave function violates both the 
total spin and the space reflection symmetries. Intuitively further progress 
can be achieved as follows: the orbitals (see Fig.\ 3) of the sS-UHF can be
viewed as optimized atomic orbitals (OAO's) to be used in constructing a
many-body wave function conforming to a Generalized Valence Bond (GVB) 
structure and exhibiting the correct symmetry properties.
Starting with the sS-UHF symmetry-breaking Slater determinant, this program
can be systematically carried out within the framework of the theory of
Restoration of Symmetry (RS) via the so-called Projection Techniques.
Instead of the more complicated case of the Li$_2$-QDM, and for 
reasons of simplicity and conceptual clarity, we will present below the case 
of the QD molecular hydrogen as an illustrative example of this RS procedure.

\newpage
\section{Artificial molecular hydrogen (H$_2$-QDM) in a Generalized Valence 
Bond Approach}
\label{sec:5}

\subsection{The sS-UHF description}

We turn now to the case of the H$_2$-QDM.
Fig.\ 4 displays the RHF and sS-UHF results for the $P=0$ case (singlet) and 
for an interdot distance $d=30$ nm and a barrier $V_b=4.95$ meV. In the RHF
(Fig.\ 4, left), both the spin-up and spin-down electrons occupy the same 
bonding ($\sigma_g$) molecular orbital. In contrast, in the sS-UHF result the 
spin-up electron occupies an optimized $1s$ (or $1s_L$) atomic-like orbital
(AO) in the left QD, while the spin down electron occupies the 
corresponding $1s^\prime$ (or $1s_R$) AO in the right QD. Concerning the total
energies, the RHF yields $E_{RHF}(P=0)=13.68$ meV, while the sS-UHF
energy is $E_{sSUHF}(P=0)=12.83$ representing a gain in energy of 0.85
meV. Since the energy of the triplet is $E_{sUHF}(P=1)=E_{sSUHF}(P=1)=
13.01$ meV, the sS-UHF singlet conforms to the requirement that
for two electrons at zero magnetic field the singlet is always the
ground state; on the other hand the RHF MO solution fails in this respect. 

\begin{figure}[t]
\vspace{3cm}
\centering{\Large{\bf FIGURE 4}}
\vspace{3cm} 
\caption{
Lateral H$_2$ QDM: Occupied orbitals (modulus square, bottom half) and total 
charge (CD) and spin (SD) densities (top half) for the $P=0$ spin unpolarized 
case. Left column: RHF results. Right column: sS-UHF results exhibiting
a breaking of the space symmetry. The numbers displayed with each orbital 
are their eigenenergies in meV, while the up and down arrows indicate an 
electron with an up or down spin. The numbers displayed with the charge 
densities are the total energies in meV. Unlike the RHF case, the spin density
of the sS-UHF exhibits a well developed spin density wave. 
Distances are in nm and the electron densities in $10^{-4}$ 
nm$^{-2}$. The choice of parameters is: $m^*=0.067 m_e$, $\hbar \omega_0=5$ 
meV, $d=30$ nm, $V_b=4.95$ meV, $\kappa=20$.
}
\end{figure}

\subsection{Projected wave function and restoration of the broken symmetry}

To make further progress, we utilize the spin projection technique to
restore the broken symmetry of the sS-UHF determinant (henceforth we will
drop the prefix sS when referring to the sS-UHF determinant),
\begin{eqnarray}
\sqrt{2}\Psi_{UHF}(1,2) &=& 
\left|
\begin{array}{cc}
u({\bf r}_1) \alpha(1) \; & \; v({\bf r}_1) \beta(1) \\
u({\bf r}_2) \alpha(2) \; & \; v({\bf r}_2) \beta(2) 
\end{array}
\right|  \nonumber \\
&\equiv& | u(1) \bar{v}(2) >~,
\label{det}
\end{eqnarray}
where $u({\bf r})$ and $v({\bf r})$ are the $1s$ (left) and $1s^\prime$ (right)
localized orbitals of the sS-UHF solution displayed in the right column of
Fig.\ 4; $\alpha$ and $\beta$ denote the up and down spins, respectively. In 
Eq.\ (\ref{det}) we also define a compact notation for the $\Psi_{UHF}$ 
determinant, where a bar over a space orbital denotes a spin-down electron;
absence of a bar denotes a spin-up electron.

$\Psi_{UHF}(1,2)$ is an eigenstate of the projection $S_z$ of the
total spin ${\bf S} = {\bf s}_1 + {\bf s}_2$, but not of $S^2$.
One can generate a many-body wave function which is an
eigenstate of $S^2$ with eigenvalue $s(s+1)$ by applying the following 
projection operator introduced by L\"{o}wdin \cite{low,paun},
\begin{equation}
P_s \equiv \prod_{s^\prime \neq s}
\frac{S^2 - s^\prime(s^\prime + 1) \hbar^2}
{[s(s+1) - s^\prime(s^\prime + 1)] \hbar^2}~,
\label{prjp}
\end{equation}
where the index $s^\prime$ runs over the eigenvalues of $S^2$.

The result of $S^2$ on any UHF determinant can be calculated with the
help of the expression,
\begin{equation}
S^2 \Phi_{UHF} = 
\hbar^2 \left[ (n_\alpha - n_\beta)^2/4 + n/2 + \sum_{i<j} 
\varpi_{ij} \right]
\Phi_{UHF}~,
\label{s2}
\end{equation}
where the operator $\varpi_{ij}$ interchanges the spins of electrons
$i$ and $j$ provided that their spins are different; $n_\alpha$ and
$n_\beta$ denote the number of spin-up and spin-down electrons, respectively,
while $n$ denotes the total number of electrons. 

For the singlet state of two electrons, $n_\alpha=n_\beta=1$,
$n=2$, and $S^2$ has only the two eigenvalues $s=0$ and $s=1$.
As a result,
\begin{eqnarray}
2 \sqrt{2} P_0 \Psi_{UHF}(1,2) & = & (1- \varpi_{12}) \sqrt{2} \Psi_{UHF}(1,2) 
\nonumber \\
& = & | u(1) \bar{v}(2) > - | \bar{u}(1) v(2) >~.
\label{prj0}
\end{eqnarray}
In contrast to the single-determinantal wave functions of the RHF and sS-UHF 
methods, the projected many-body wave function (\ref{prj0}) is a linear
superposition of two Slater determinants, and thus represents a corrective step
beyond the mean-field approximation.

Expanding the determinants in Eq.\  (\ref{prj0}), one finds the equivalent
expression
\begin{equation}
2 P_0 \Psi_{UHF}(1,2)  = 
(u({\bf r}_1)v({\bf r}_2)+u({\bf r}_2)v({\bf r}_1)) \chi(0,0)~,
\label{hl1}
\end{equation}
where the spin eigenfunction for the singlet is given by
\begin{equation}
\chi(s=0,S_z=0)=(\alpha(1)\beta(2)-\alpha(2)\beta(1))/\sqrt{2}~.
\label{spin0}
\end{equation}
Eq.\ (\ref{hl1}) has the form of a Heitler-London (HL) \cite{hel} 
or valence bond \cite{coul,murr,note2} wave function.
However, unlike the HL scheme which uses the orbitals $\phi_L({\bf r})$ and
$\phi_R({\bf r})$ of the separated (left and right) atoms \cite{note3}, 
expression (\ref{hl1}) employs
the sS-UHF orbitals which are self-consistently optimized for any separation 
$d$ and potential barrier height $V_b$. As a result, expression (\ref{hl1}) 
can be characterized as a Generalized Valence Bond (GVB) \cite{note4}
wave function. Taking into account the normalization of the spatial part, we 
arrive at the following improved wave function for the singlet state 
exhibiting all the symmetries of the original many-body hamiltonian,
\begin{equation}
\Psi^{\rm s}_{GVB}(1,2) = N_+ \sqrt{2} P_0 \Psi_{UHF}(1,2)~,
\label{gvb}
\end{equation}
where the normalization constant is given by
\begin{equation}
N_+ = 1/\sqrt{1+S^2_{uv}}~,
\end{equation}
$S_{uv}$ being the overlap integral of the $u({\bf r})$ and $v({\bf r})$ 
orbitals,
\begin{equation}
S_{uv}= \int u({\bf r})v({\bf r}) d{\bf r}~.
\end{equation}

The total energy of the GVB state is given by
\begin{equation}
E^{\rm s}_{GVB}=N_+^2 [h_{uu}+h_{vv}+2S_{uv}h_{uv}+J_{uv}+K_{uv}]~,
\label{engvb}
\end{equation}
where $h$ is the single-particle part of the total hamiltonian (\ref{hsp}),
and $J$ and $K$ are the direct and exchange matrix elements associated
with the $e-e$ repulsion $e^2/\kappa r_{12}$. For comparison, we give also
here the corresponding expression for the HF total energy either in 
the RHF ($v=u$) or sS-UHF case, 
\begin{equation}
E^{\rm s}_{HF}=h_{uu}+h_{vv}+J_{uv}~.
\label{enghf}
\end{equation}

For the triplet, the projected wave function coincides with the original
HF determinant, so that the corresponding energies in all three 
approximation levels are equal, i.e., $E^{\rm t}_{GVB}=E^{\rm t}_{RHF}=
E^{\rm t}_{UHF}$.
 
\begin{figure}[t]
\centering\includegraphics[width=8.2cm]{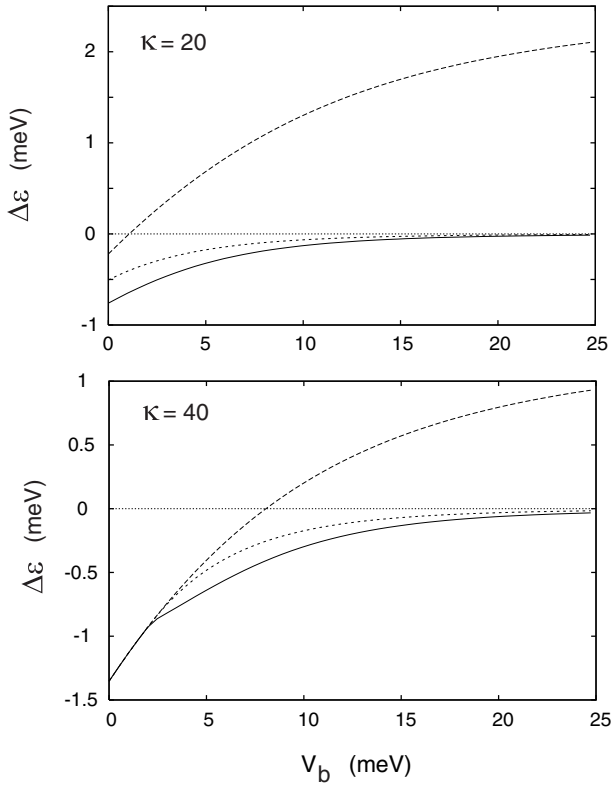}\\
~~~~\\
\caption{
Lateral H$_2$-QDM: The energy difference between the singlet and
triplet states according to the RHF (MO Theory, top line), the sS-UHF
(middle line), and the GVB approach (Projection Method, bottom line) as a 
function of the interdot barrier
$V_b$. For $V_b=25$ meV complete dissociation has been clearly reached.
Top frame: $\kappa=20$. Bottom frame: $\kappa=40$.
The choice of the remaining parameters is: $m^*=0.067 m_e$, 
$\hbar \omega_0=5$ meV and $d=30$ nm.  
}
\end{figure}
\subsection{Comparison of RHF, sS-UHF and GVB results}

A major test for the suitability of different methods to the case of
QDM's is their ability to properly describe the dissociation limit of
the molecule. The H$_2$-QDM dissociates into two non-interacting
QD hydrogen atoms with arbitrary spin orientation. As a result, the
energy difference, $\Delta \varepsilon=E^{\rm s}-E^{\rm t}$, between the 
singlet and the triplet states of the molecule must approach the zero
value from below as the molecule dissociates. To theoretically generate
such a dissociation process, we keep the separation $d$ constant and
vary the height of the interdot barrier $V_b$; an increase in the value of
$V_b$ reduces the coupling between the individual dots and for sufficiently 
high values we can always reach complete dissociation.  

Fig.\ 5 displays the evolution in zero magnetic field of $\Delta \varepsilon$ 
as a function of $V_b$ and for all three approximation levels, i.e., the RHF
(MO Theory, top line), the sS-UHF (middle line), and the GVB (bottom line). 
The interdot distance is the same as in Fig.\ 4, i.e., $d=30$; the case of
$\kappa=20$ is shown at the top panel, while the case of a weaker
$e-e$ repulsion is displayed for $\kappa=40$ at the bottom panel.

We observe first that the MO approach fails completely to describe
the dissociation of the H$_2$-QDM, since it predicts a strongly stabilized 
ferromagnetic ordering in contradiction to the expected singlet-triplet 
degeneracy upon full separation of the individual dots.
A second observation is that both the sS-UHF and the GVB solutions
describe the dissociation limit ($\Delta \varepsilon \rightarrow 0$ for
$V_b \rightarrow \infty$) rather well. In particular in both the sS-UHF
and the GVB methods the singlet state remains the ground state for all
values of the interdot barrier. Between the two singlets, the GVB one
is always the lowest, and as a result, the GVB method presents an
improvement over the sS-UHF method both at the level of symmetry
preservation and the level of energetics.

It is interesting to note that for $\kappa=40$ (weaker $e-e$ repulsion)
the sS-UHF and GVB solutions collapse to the MO solution for smaller
interdot-barrier values $V_b \leq 2.8$ meV. However, for the stronger
$e-e$ repulsion ($\kappa=20$) the sS-UHF and GVB solutions remain energetically
well below the MO solution even for $V_b=0$. Since the separation
considered here ($d=30$ nm) is a rather moderate one (compared to the value 
$l_0=28.50$ nm for the extent of the $1s$ AO), we conclude that there is 
a large range of materials parameters and interdot distances for which
the QDM's are weakly coupled and cannot be described by the MO theory.

\section{Conclusions}
\label{sec:6}

We have shown that the sS-UHF method, in conjunction with the companion
step of the restoration of symmetries when such symmetries are broken, is
able to describe the full range of couplings in a QDM, from the
strong-coupling regime exhibiting delocalized molecular orbitals to the 
weak-coupling one associated with Heitler-London-type combinations
of atomic orbitals.

The breaking of space symmetry within the sS-UHF
method is necessary in order to properly describe the weak-coupling
and dissociation regimes of QDM's. The breaking of the space symmetry
produces optimized atomic-like orbitals localized on each individual dot. 
Further improvement is achieved with the help of Projection Techniques 
which restore the broken symmetries
and yield multideterminantal many-body wave functions. The method of the 
restoration of symmetry was explicitly illustrated for the case of the 
molecular hydrogen QD (H$_2$-QDM, see section 5). It led to the introduction 
of a Generalized Valence Bond many-body wave function as the 
appropriate vehicle for the description of the weak-coupling and dissociation
regimes of artificial molecules. In all instances the weak-coupling
regime is accompanied by an antiferromagnetic ordering of the
spins of the individual dots.

Additionally, we showed that the RHF, whose orbitals  preserve the space 
symmetries and are delocalized over the whole molecule, is naturally
associated with the molecular orbital theory. As a result, and in
analogy \cite{coul,murr,szab} with the natural molecules, it was found 
that the RHF fails to describe the weak-coupling and dissociation regimes of
QDM's. It can further be concluded that the spontaneous polarization
and ferromagnetic ordering predicted for weakly-coupled double QD's in Refs.\
\cite{nag,wen} is an artifact of the molecular-orbital structure implicit
\cite{note5} in the framework of LSD calculations.


\begin{thebibliography}{99}
\bibitem{hf}
Within the terminology adopted here, the simple designation Hartree-Fock (HF) 
in the literature most often refers to our restricted HF (RHF), in particular
in atomic physics and the physics of the homogeneous electron gas. In nuclear
physics, however, the simple designation HF most often refers to a space
(S)-UHF. The simple designation unrestricted Hartree-Fock (UHF) in Chemistry
most often refers to our s-UHF.
\bibitem{note12}
The parameter controlling the appearance of electron localization within
each individual dot is $R_W \equiv Q/\hbar \omega_0$, where $Q$ is the
Coulomb interaction strength and $\hbar \omega_0$ is the characteristic 
energy of the parabolic
confinement; $Q=e^2/\kappa l_0$, with $l_0=(\hbar/m^* \omega_0)^{1/2}$ 
being the spatial extent of the lowest state's wave function in the parabolic 
confinement. With the sS-UHF, we find that Wigner crystallization occurs 
in both the single QD's and QDM's for $R_W > 1$. For $\kappa=20$, one
has $R_W=0.95$.
\end{thebibliography}
\end{document}